\documentclass[prl,reprint,amsmath,amssymb,aps]{revtex4-2}
\usepackage{dcolumn}
\usepackage{bbm}
\usepackage{subcaption}

\usepackage{graphicx}
\usepackage{color}
\usepackage{dsfont}

\usepackage{bm}

\usepackage{enumitem}

\begin{document}
\title{Black Hole States in Quantum Spin Chains}

\author{Charlotte Kristjansen$^a$ and Konstantin Zarembo$^{a,b}$}
\affiliation{$^a$Niels Bohr Institute, Copenhagen University, Copenhagen, Denmark}
\affiliation{$^b$Nordita, KTH Royal Institute of Technology and Stockholm University,
Stockholm, Sweden}

\begin{abstract}
We define a black hole state in a spin chain by studying thermal correlators in holography. Focusing on the Heisenberg model we investigate the thermal and complexity properties of the black hole state by evaluating its entanglement entropy, emptiness formation probability and Krylov complexity. The entanglement entropy grows logarithmically with effective central charge $c\simeq 5.2$. We also find evidence for thermalization at infinite temperature.
\end{abstract}

\maketitle
\section{The Black Hole State \label{Intro}}

At the heart of the gauge–gravity duality lies a remarkable identification of a thermal ensemble in field theory with a black hole in the dual gravitational description \cite{Witten:1998zw}. The infall of a particle into the black hole permits local operators to exist in isolation and  gives rise to  their thermal one-point functions which thereby serve as sensitive probes of the black-hole interior \cite{Grinberg:2020fdj}. 

In a more refined holographic picture particles are replaced by strings, which in turn admit an effective spin-chain description. The key advantage of the latter is manifest integrability  \cite{Minahan:2002ve}. The  one-point functions  (in other setups where they arise) map to overlaps of the spin-chain boundary states and can be efficiently studied by integrability methods \cite{deLeeuw:2015hxa,Buhl-Mortensen:2015gfd} (see \cite{Kristjansen:2024dnm,Linardopoulos:2025ypq} for a review). Since black holes are chaotic systems, we expect the integrability to be broken in the thermal ensemble, and we will later confirm this expectation. 

The resulting boundary state nonetheless exhibits remarkable features which we believe are of  interest on their own. Our ultimate goal is to study boundary states that arise in AdS/CFT, but this requires  extra formalism while  succinct features can be illustrated in a simpler setting of a spin-1/2 Heisenberg model:
\begin{equation}\label{Hamilton}
 H=\sum_{\ell=1}^{L}\left(\bm{\sigma }_{\ell}\cdot \bm{\sigma }_{\ell+1}
 +\lambda\, \bm{\sigma }_{\ell}\cdot \bm{\sigma }_{\ell+2}\right).
\end{equation}
We will construct what we call {\it the black-hole states} in this model using simple plausibility arguments and later show that their direct counterparts govern thermal one-point functions in AdS/CFT. Other approaches to defining black hole states  based on discrete models and 
thermal correlation functions can be found in~\cite{Janik:2025zji}, where the black hole emerges from the 2D Ising model, and in~\cite{Basteiro:2022zur,Basteiro:2024crz,Dey:2024jno} where it occurs through a discretization of hyperbolic space. 

Since the spin chain of AdS/CFT is integrable, we mostly concentrate on the integrable case without the next-to-nearest interactions. Non-zero $\lambda $ breaks integrability making the model fully chaotic for $\lambda \gtrsim 0.5$. We will use this to see how much thermalization of the balck-hole state depends on the chaotic nature of the Hamiltonian. 

\begin{figure}[t]
  \centering
  \includegraphics[width=1\linewidth]{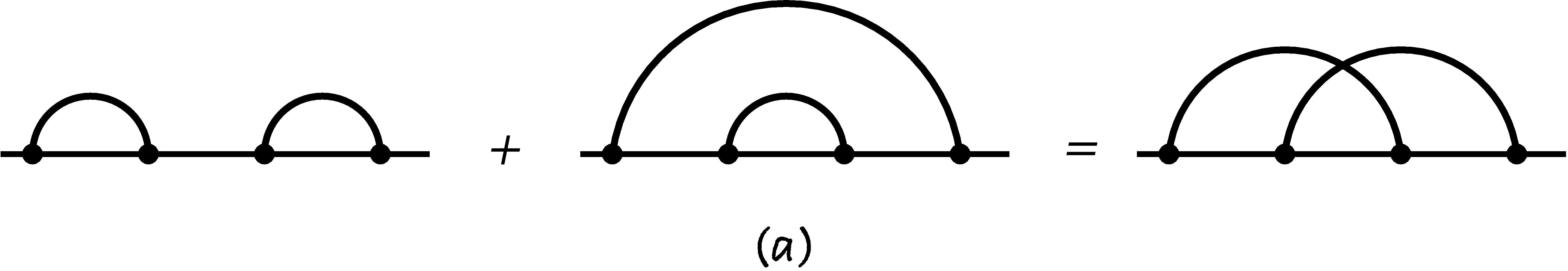}\hfill
  \vspace*{0.6cm}
  \includegraphics[width=1\linewidth]{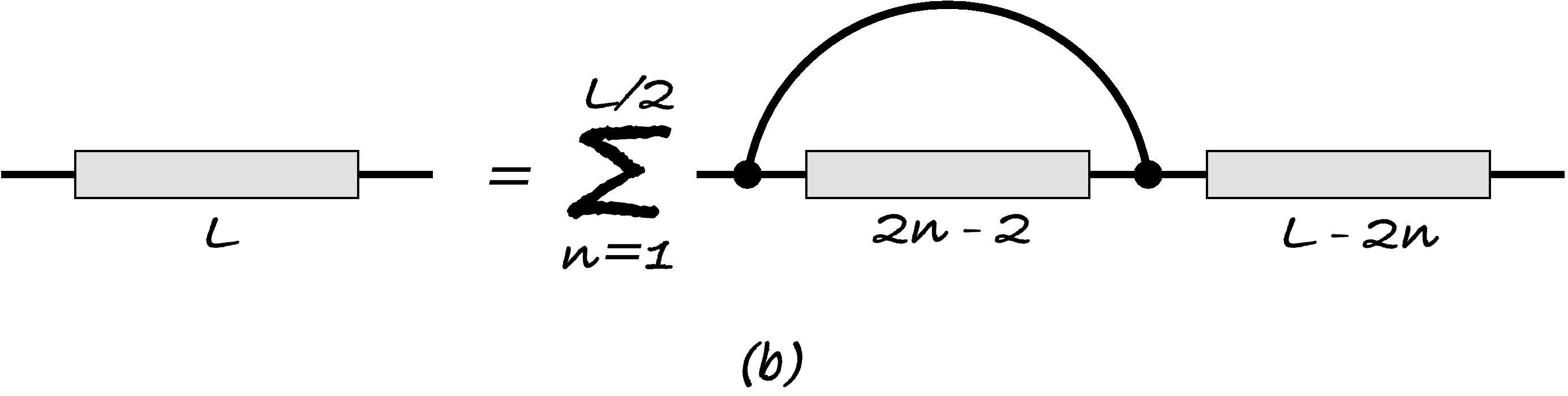}
  \caption{\small (a) Rumer relation.
           (b) Recursion relation defining the black-hole states.
         }
  \label{fig:multifig}
\end{figure}

The black-hole state should be a spin singlet (black holes have no hair) and it should be sufficiently generic to reflect black hole's chaotic nature. In order to systematically enumerate singlets 
we invoke well-known results from the quantum-mechanical theory of valence, originally due to Rumer \cite{rumer1932theorie,weyl1932valenztheorie}. 

The singlet formed by two spins is their anti-symmetric combination: $\varepsilon _{\alpha _1\alpha _2}$, where $\alpha_i\in \{ \uparrow,\downarrow\}$. Four spins can be split in pairs in three possible ways,
 but only two of them are linearly independent, because of the identity  $\varepsilon _{\alpha _1\alpha _2}\varepsilon _{\alpha _3\alpha _4}+\varepsilon _{\alpha _1\alpha _4}\varepsilon _{\alpha _2\alpha _3}=\varepsilon _{\alpha _1\alpha _3}\varepsilon _{\alpha _2\alpha _4}$   \cite{rumer1932theorie} illustrated in fig.\ref{fig:multifig}a. Rumer's identity can be used to eliminate crossings for any number of spins. The non-crossing ("kreuzungslose") pairings are linearly independent and form a complete albeit non-orthogonal basis   \cite{weyl1932valenztheorie}. The number of pairings, for $L$ spins, is the Catalan number $C_{L/2}$, and this is precisely the dimension of the singlet subspace for the spin chain of length $L$ \cite{hulthen1938}.

 We define the black-hole state as the sum of all non-crossing pairings taken with equal weights. The state can be built recursively from the relation
 \begin{equation}\label{eq:BHstate}
 \left|{\rm BH}_L\right\rangle=\sum_{n=1}^{L/2}\sum_{\alpha \beta }^{}
 \varepsilon _{\alpha \beta }\left|\alpha \right\rangle
 \otimes\left|{\rm BH}_{2n-2}\right\rangle
 \otimes\left|\beta \right\rangle
 \otimes\left|{\rm BH}_{L-2n}\right\rangle,
\end{equation}
 as illustrated in
  fig.~\ref{fig:multifig}b. Catalan numbers, to mention in passing, satisfy pictorially equivalent recursion.
 
The non-crossing pairings are identical to t'~Hooft's planar diagrams, and the wavefunction of the black-hole state can be alternatively computed from a matrix integral:
\begin{equation}\label{MM-def-ferm}
 {\rm BH}^{\alpha _1\ldots \alpha _L}=
 \lim_{N\rightarrow \infty }
 \int_{}^{}d\Psi ^\uparrow \,d\Psi ^\downarrow\,
 \,{\rm e}\,^{-\mathop{\mathrm{tr}}\bar{\Psi }\Psi }
 \mathop{\mathrm{tr}}
 \Psi ^{\alpha _1}\ldots \Psi ^{\alpha _L}.
 \end{equation}
 The integration here is over two $N\times N$ Hermitian matrices with Grassmann entries. The bar denotes Majorana conjugation:
 \begin{equation}
 \bar{\Psi }_\beta  =\Psi ^\alpha  \varepsilon _{\alpha \beta }.
\end{equation}
In the exponent therefore stands  the standard action for a zero-dimensional Majorana fermion. Integration has to be Grassmann, bosonic matrices would be incompatible with the cyclicity of the trace and anti-symmetry of $\varepsilon _{\alpha \beta }$.
The Wick contractions among $\Psi ^{\alpha _\ell}$ produce all possible spin pairings and in the large-$N$ limit only planar contractions survive. The recursion relation is simply the Schwinger-Dyson equation of the matrix integral.

The black-hole state is manifestly $SU(2)$-invariant,  changes sign under cyclic permutations (has momentum $\pi $), and covers evenly the singlet subspace of the spin chain's Hilbert space. In some sense it is just the familiar matrix product state but with bond matrices drawn at random from the Gaussian ensemble. Random matrix product states have been introduced recently in a quite similar context \cite{Jung:2025rip}.

The  Heisenberg model is integrable, its Hamiltonian commutes with an infinite number of  extra conserved charges starting with $Q_3=i\sum_{\ell}^{}\bm{\sigma }_\ell\cdot [\bm{\sigma }_{\ell+1}\times \bm{\sigma }_{\ell+2}]$. A boundary state is compatible with this  structure  if annihilated by $Q_3$ (and higher odd charges as well) \cite{Ghoshal:1993tm,Piroli:2017sei}. Integrable boundary states have remarkable mathematical properties, in particular, their overlaps with the physical eigenstates of the Hamiltonian can be systematically computed by  Bethe ansatz \cite{Brockmann:2014a,Gombor:2023bez,Gombor:2024iix}.
The black-hole state however breaks integrability. It is easy to check that $Q_3\left|{\rm BH}\right\rangle\neq 0$ at length six and higher.  We therefore expect that the black hole state exhibits some chaotic features even if the underlying Hamiltonian is completely integrable.

We will probe the black-hole state with the standard diagnostics of quantum chaos: entanglement entropy, Krylov complexity and eigenvalue thermalization. 
For comparison, we also consider a singlet valence-bond state (VBS) which is known to preserve integrability \cite{Pozsgay:2018ixm}: 
\begin{equation}\label{VBS}
 {\rm VBS}^{\alpha _1\ldots \alpha _L}=\varepsilon ^{\alpha _1\alpha _2}\ldots \varepsilon ^{\alpha _{L-1}\alpha _L}.
\end{equation}
It has the same quantum numbers  as the black-hole state upon averaging over translations by one site: 
 $\left\langle {\rm VBS}_\pi \right|=\left\langle {\rm VBS}\right|(1-T)$. As we shall see it has very different entanglement properties and much lower Krylov complexity.

\section{Entanglement Entropy\label{EEntropy}}
The Von Neumann entanglement entropy can be defined for a state of a quantum system whose Hilbert space factorizes
into a tensor product:
\begin{equation}
{\cal{H}}={\cal{H}}_A \otimes {\cal{H}}_B.
\end{equation}
If the state of the total quantum system is described by the density matrix $\rho$ on $\cal H$, one defines the reduced
density matrix corresponding to the subspace ${\cal H}_{A}$ as
\begin{equation}
\rho_{ A}=\mbox{Tr}_{B}\rho.
\end{equation}
The entanglement entropy of subsystem $A$ with respect to $B$ is then
\begin{equation}
S(A)=-\mbox{Tr}\rho_A \log \rho_A.
\end{equation}
Even if the total system is in a pure state $\rho =\left|\psi \right\rangle\left\langle \psi \right|$, entanglement between subsystems generates an entropy.

A simple recipe for computing the entanglement entropy for a (normalized) spin chain state, $|\psi \rangle$, is to
express the state in a product basis as follows, see e.g. \cite{Nielsen:2012yss}
\begin{equation}
|\psi\rangle = \sum_{i,j} M_{ij}  \,|i\rangle_A \, |j\rangle_B,
\end{equation}
where the $|i\rangle_A$ and the $|j\rangle_B$ constitute an orthonormal basis of respectively the subspace $A$ and the subspace $B$.  Then it holds that
\begin{equation}
\rho_A= M M^\dagger,
\end{equation}
and 
\begin{equation}
S(A)=-\sum_k \lambda_k \log\lambda_k,
\end{equation}
for the non-vanishing eigenvalues $\lambda_k$ of $M M^{\dagger}$.

We will consider the entanglement entropy of a subspace  of the spin chain of length $\ell$ relative to the full chain 
of length $L$, denoted in the following as $S(\ell)$, for respectively the black hole state and the integrable valence-bond state.

\begin{figure}[t]
 \centerline{\includegraphics[width=8cm]{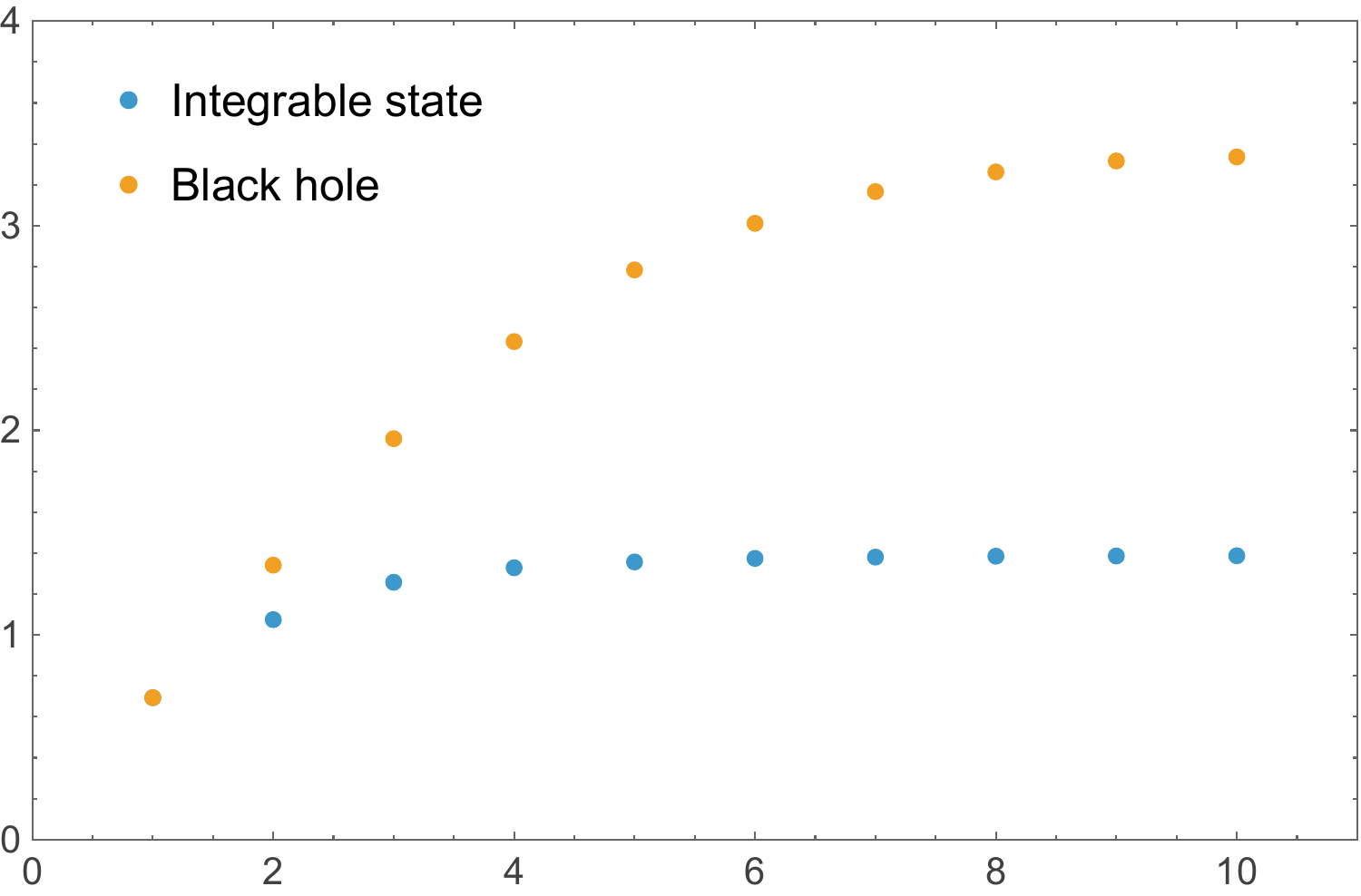}}
\caption{\label{EE} \small The entanglement entropy as a function of $\ell$ for the integrable and
black hole states at $L=20$.}
\end{figure}

In a critical (gapless) system, the entanglement entropy of the ground state
can be computed by 2d CFT methods and
for periodic boundary conditions is
expected to behave as~\cite{Calabrese:2004eu}  
\begin{equation}
S(\ell)=\frac{c}{3}\log\left(\frac{L}{\pi} \sin \frac{\pi \ell}{L} \right) +c_1', \label{eescaling}
\end{equation}
where $c$ is the central charge of the CFT and $c_1'$ is a non-universal numerical constant. This is in contrast to a gapped ground
state which will have constant entanglement entropy and thermal states for which the entanglement entropy grows
linearly with $\ell$~\cite{Laflorencie:2015eck}.
The behavior in~(\ref{eescaling}) was confirmed for the ground state of the critical Heisenberg XXX$_{1/2}$ chain~\cite{Latorre:2003kg,Korepin:2004zz,Affleck:1986bv}.
The central charge in this case equals one: $c=1$.

In figure~\ref{EE} we show the entanglement entropy $S(\ell)$ as a function of $\ell$ for respectively the integrable and
the black hole state for L=20. We remark that $S(\ell)$ for $l=1$ is always equal to $\log 2$ for the states that we consider due
to their singlet nature. Furthermore, we notice that the integrable two site singlet product state has an entanglement entropy which saturates at $\log 4$ originating from the fact that only two singlet bonds are broken when we cut
out an interval of length $\ell$. In contrast the entropy of the black-hole state continues to grow in almost perfect agreement with the CFT prediction~(\ref{eescaling}). 

\begin{figure}[t]
 \centerline{\includegraphics[width=8cm]{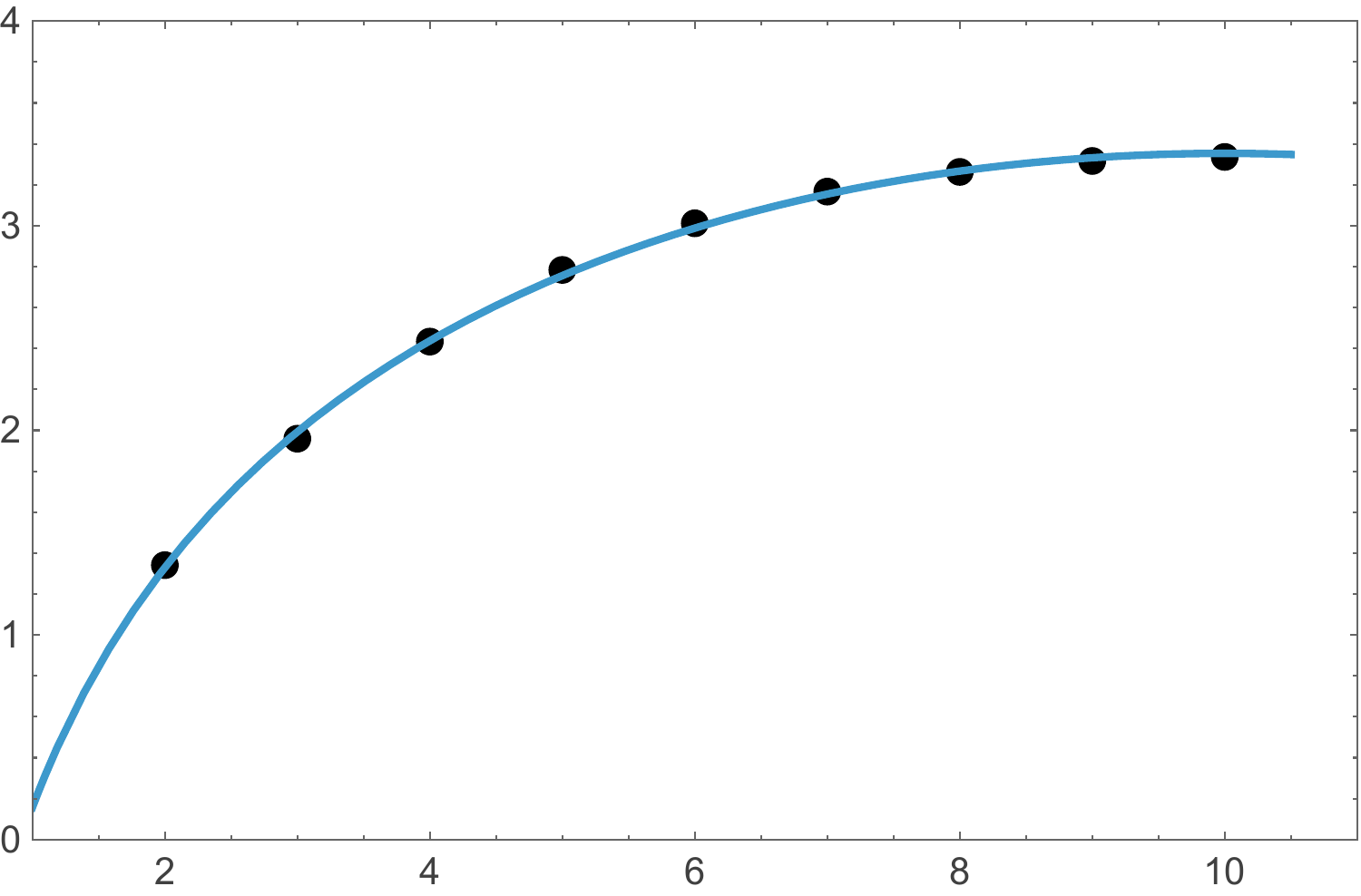}}
\caption{\label{EEfitted} \small The von Neumann entanglement entropy as a function of $\ell$ for the
black hole state for $L=20$ and the corresponding fit to the functional form given in eqn.~(\ref{eescaling}).}
\end{figure}

In figure~\ref{EEfitted} we fit the entanglement entropy for the black hole state to the functional form of eqn.~(\ref{eescaling}). The fit is remarkably accurate and produces the following values
for $c$ and $c_1'$:
\begin{equation}
c=5.2,\hspace{1.cm} c_1'=0.164.
\end{equation}
The physical interpretation of this result is unclear to us. We do not know if it is meaningful to assign a CFT to a state, and what this CFT can be.

\section{Thermalization \label{Thermalization}}

One possible formulation of the Eigenvalue Thermalization Hypothesis  (ETH) posits that time averages in a sufficiently generic state thermalize \cite{Rigol:2012yjf,DAlessio:2015qtq,Alishahiha:2025rdg}: 
\begin{equation}
 \overline{\left\langle \Psi \right|A(t)\left|\Psi \right\rangle}
 =\mathop{\mathrm{tr}}\rho A,
 \qquad 
\rho =\frac{1}{Z}\,\,{\rm e}\,^{-\beta H}.
\end{equation}
The time average, in absence of accidental degeneracies, reduces to the statistical average in the diagonal ensemble:
\begin{equation}\label{diag-ens}
  \overline{\left\langle \Psi \right|A(t)\left|\Psi \right\rangle}
  =\sum_{n}^{}\left|\left\langle \Psi \right.\!\left|n \right\rangle\right|^2
  \left\langle n\right|A\left|n\right\rangle.
\end{equation}
Of course the diagonal ensemble does  not literally coincide with the grand-canonical one, but
course-graining in the infinite-volume limit converts, it is believed, one to the other provided the initial state $\left|\Psi \right\rangle$ is sufficiently generic  \cite{rigol2008thermalization}. The grand-canonical ensemble of an integrable model contains infinitely many chemical potentials, one for each conserved charge, and ETH does not imply such a remarkable reduction in the degrees of freedom as for chaotic systems. 
 Surprisingly, the black-hole state seems to thermalize without any dependence on the higher conserved charges, not even on the Hamiltonian itself, and moreover this happens for integrable as well as non-integrable Hamiltonians.

We will compare the diagonal ensemble with the infinite-temperature one. The density matrix of the latter is the projector onto the singlet subspace:
\begin{equation}\label{beta=0-ensemble}
 \rho_\infty  =\frac{1}{C_{L/2}}\,\Pi _{\rm s},
 \qquad 
 \Pi _{\rm s} =\int_{SU(2)}^{}dg\,g\otimes\ldots \otimes g,
\end{equation}
where $dg$ is the Haar measure on the $SU(2)$ group manifold. The black-hole state obviously cannot cover the whole Hilbert space of the spin chain, having overlap only with singlet eigenstates.

To probe the diagonal ensemble we study the emptiness formation probability (EFP) \cite{Korepin:1994ui}:
\begin{equation}
 P(n)=\prod_{\ell=1}^{n}\frac{1+\sigma ^3_\ell}{2}\,,
\end{equation}
that projects  $n$ consecutive spins onto the up state. The expectation value of EFP
decays exponentially with the length of the string: $\left\langle P(n)\right\rangle\sim \,{\rm e}\,^{-f n}$ with the exponent given by the free energy density 
\cite{Boos:2001cd,Boos:2002gu,Abanov:2002wm}. The latter is 
temperature-dependent, making EFP an ideal probe of thermalization.

\begin{figure}[t]
  \centering
  \begin{subfigure}{0.9\linewidth}
    \includegraphics[width=\linewidth]{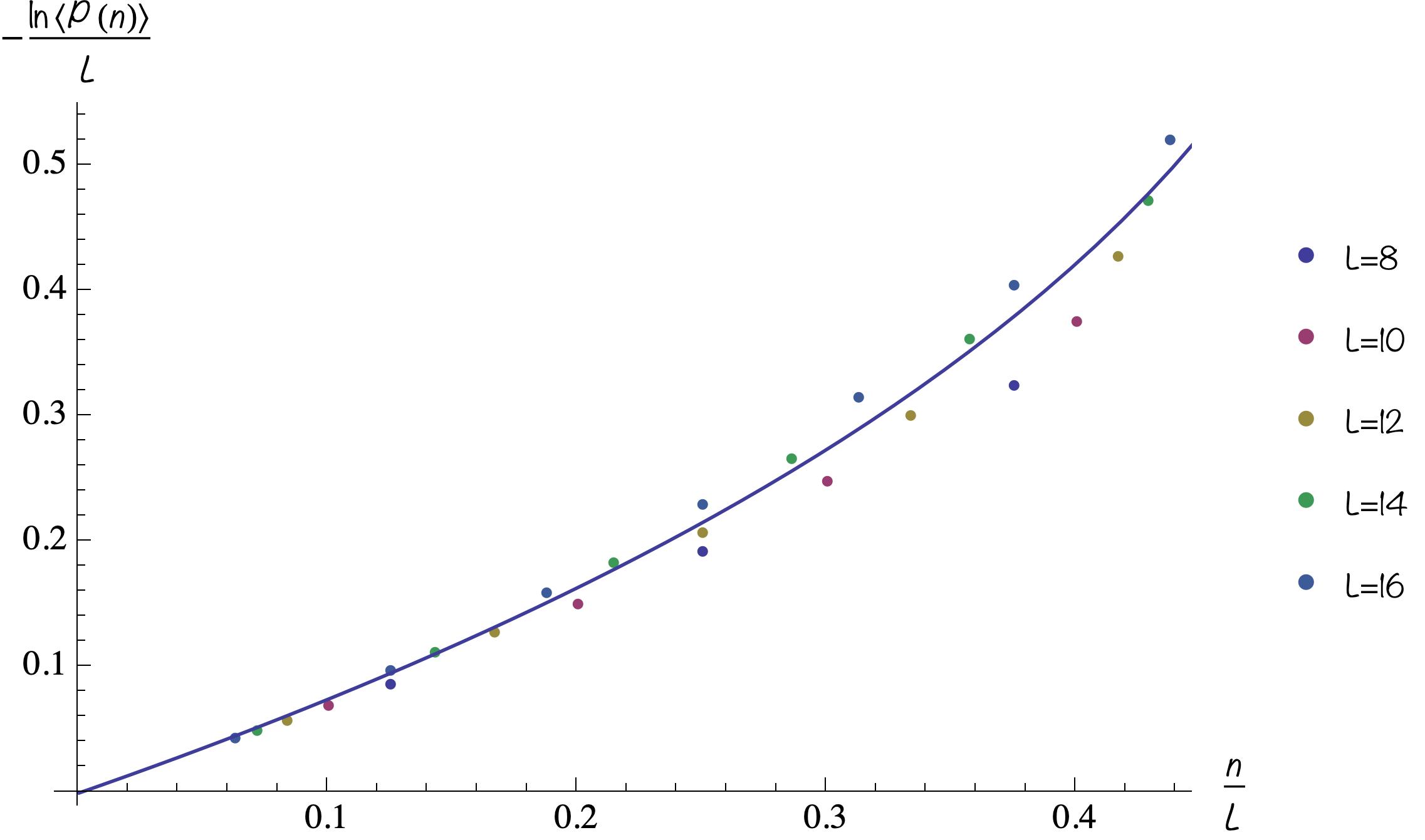}
    \label{fig:1}
  \end{subfigure}
  \hfill
  \begin{subfigure}{0.9\linewidth}
    \includegraphics[width=\linewidth]{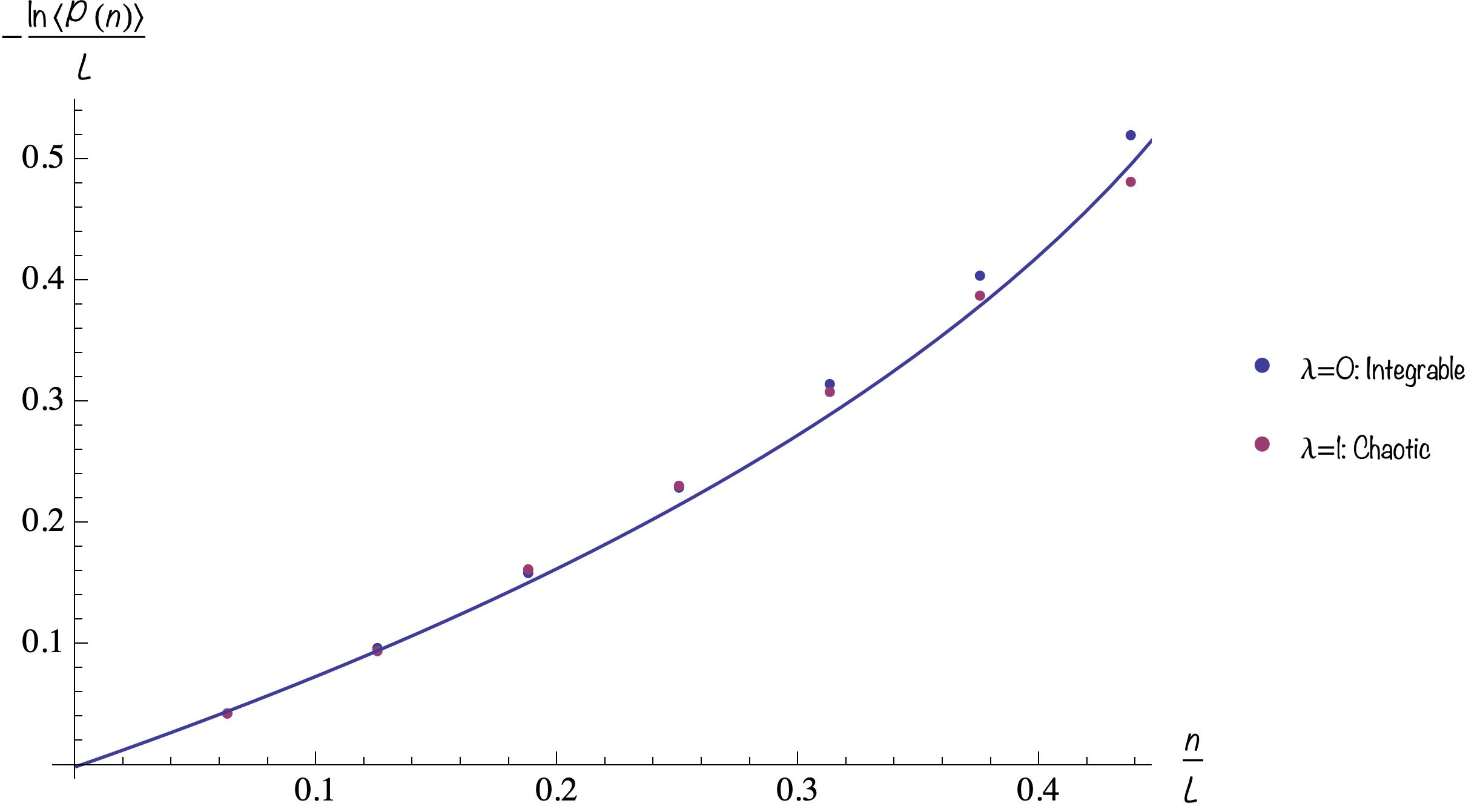}
    \label{fig:2}
  \end{subfigure}
  \caption{\label{emptiness-figure}\small The emptiness formation probability in the diagonal ensemble with Boltzmann weights $\left\langle {\rm BH}\right.\!\left|n \right\rangle^2$, compared to the statistical average at infinite temperature: the solid line is (\ref{th-exponent}). The upper plot is for the Heisenberg Hamiltonian, $\lambda =0$ in (\ref{Hamilton}). In the lower plot the Heisenberg model  is compared to a non-integrable case with $\lambda =1$ for the spin chain of length $16$.}
\end{figure}

It is easy to calculate the EFP in the infinite-temperature ensemble (\ref{beta=0-ensemble}). In the parametrization $g=n_0+i\sigma ^in_i$ with a unit four-vector $n_\mu $ the group integrals become averages over the round three-dimensional sphere, for instance:
\begin{equation}
 \mathop{\mathrm{tr}}\Pi _{\rm s}
 =\frac{1}{2\pi ^2}\int_{}^{}d^3n\,(2n_0)^L=C_{L/2},
\end{equation}
reconfirming that the total number of singlets is  the Catalan number. Here we have taken into account that $\mathop{\mathrm{tr}}g=2n_0$. By the same token, 
 $\mathop{\mathrm{tr}}g(1+\sigma ^3)/2=n_0+in_3$, and the EFP in the infinite-temperature ensemble is
\begin{align}
 \left\langle P(n)\right\rangle_\infty &=
 \frac{1}{2\pi ^2C_{L/2}}\int_{}^{}d^3n\,(2n_0)^{L-n}(n_0+in_3)^n
 \\
 &=\frac{(L-n)!(L/2)!}{L!(L/2-n)!}\stackrel{L\rightarrow \infty }{\simeq }
 \,{\rm e}\,^{-f(n/L)L}
\end{align}
with
\begin{equation}\label{th-exponent}
 f(x)=x\ln 2+\left(\frac{1}{2}-x\right)\ln(1-2x)-(1-x)\ln(1-x).
\end{equation}

In fig.~\ref{emptiness-figure} we compare this prediction with the numerical results for the diagonal ensemble of the black-hole state. The agreement is reasonably good, given relatively small length of the spin chain. Interestingly, the results do not differ much for integrable and non-integrable Hamiltonians: in the lower plot we compare the models with $\lambda =0$ and $\lambda =1$. The former is integrable while the latter is fully chaotic by the standard level-repulsion test, nonetheless the difference is minuscule, within the accuracy set by finite-size effects.

\section{Krylov Complexity \label{Krylov}}

Krylov complexity  measures to what extent an operator or a state explores a system's Hilbert space under time evolution, i.e.\ when repeatedly acted upon by the Hamiltonian ~\cite{Parker:2018yvk,Balasubramanian:2022tpr}. An extensive recent review of the virtues and applicability of the
concept can be found in~\cite{Nandy:2024evd,Rabinovici:2025otw}. The Krylov complexity of a state, also known as the spread complexity, depends both on the Hamiltonian and on the state itself~\cite{Rabinovici:2022beu,Avdoshkin:2022xuw,Scialchi:2023bmw,Craps:2024suj,Erdmenger:2023wjg,Rabinovici:2025otw,Balasubramanian:2025xkj}.
We will be interested in studying the Krylov complexity of the  black hole state~(\ref{eq:BHstate}). For comparison we also calculated the Krylov complexity of the integrable valence bond state~(\ref{VBS}). Krylov complexity of coherent states of relevance for the semi-classical limit of 
AdS/CFT integrable models was discussed in~\cite{Das:2024tnw}.  Furthermore, holographic Krylov complexity was argued to be directly extractable from string motion
in AdS~\cite{Fatemiabhari:2025cyy}.

Starting from a given state $|\psi_0\rangle$ one defines the Krylov  space of order $K$ as the space spanned by the states
$\{ |\Psi_0\rangle, H |\Psi_0\rangle, H^2|\Psi_0\rangle, \ldots, H^{K-1}|\Psi_0\rangle\}$. Using the Lanczos algorithm
one builds from this set  an orthonormal basis $\{|\psi_0\rangle,|\psi_1\rangle,\ldots, |\psi_{K-1}\rangle\}$, 
and in the same process generates an expression for the Hamiltonian, in terms of 
Lanczos coefficients:
\begin{equation}
H=\begin{pmatrix}
a_0 & b_1 &  & & & &  \\
b_1 & a_1 & b_2 & & & &  \\
& b_2 & a_2 & b_3 & & &  \\
&  & b_3 & a_3 & \ddots & &  \\
& & & \ddots & \ddots & b_{K-1}& \\
& & &  & b_{K-1}& a_{K-1} 
\end{pmatrix}.
\end{equation}

Subsequently one studies the time evolution of  $|\psi_0\rangle $ in this new basis:
\begin{equation}
|\psi(t)\rangle = \sum_{n=0}^{K-1} \psi_n(t) |\psi_n\rangle, \hspace{0.5cm}  |\psi(0)\rangle =|\psi_0\rangle.
\end{equation}
The Schr\"{o}dinger equation for $|\psi(t)\rangle$ amounts to the following set of coupled differential
equations for the functions $\psi_n(t)$.
\begin{equation}
i\frac{\partial}{\partial t} \psi_n(t)= b_n \psi_{n-1}(t)+ a_n \psi_n(t) +b_{n+1}\psi_{n+1}(t),
\end{equation}
with the understanding that $\psi_{-1}(t)=\psi_{K}(t)=0$ and with the initial condition
\begin{equation}
\left.\psi_n(t)\right|_{t=0} = \delta_{n,0}.
\end{equation}

Introducing a Krylov space position operator $\hat{n}$ as
\begin{equation}
\hat{n}= \sum_{n=1}^{K-1} |\psi_n\rangle \langle \psi_n|,
\end{equation}
the Krylov complexity (or the spread complexity) of the initial state is defined as
\begin{equation}
C_K(t)=\langle \psi(t)| \hat{n} |\psi(t) \rangle = \sum_{n=0}^{K-1} n |\psi_n(t)|^2.
\end{equation}
A more elaborate version can be found in
\cite{Balasubramanian:2022tpr}. Time averaging is often invoked to
get rid of short-scale oscillations:
\begin{equation}
\overline{C_K}(t)= \int_0^t  C_K(s) ds /t.
\end{equation}

\begin{figure}[h]
{\includegraphics[width=8cm]{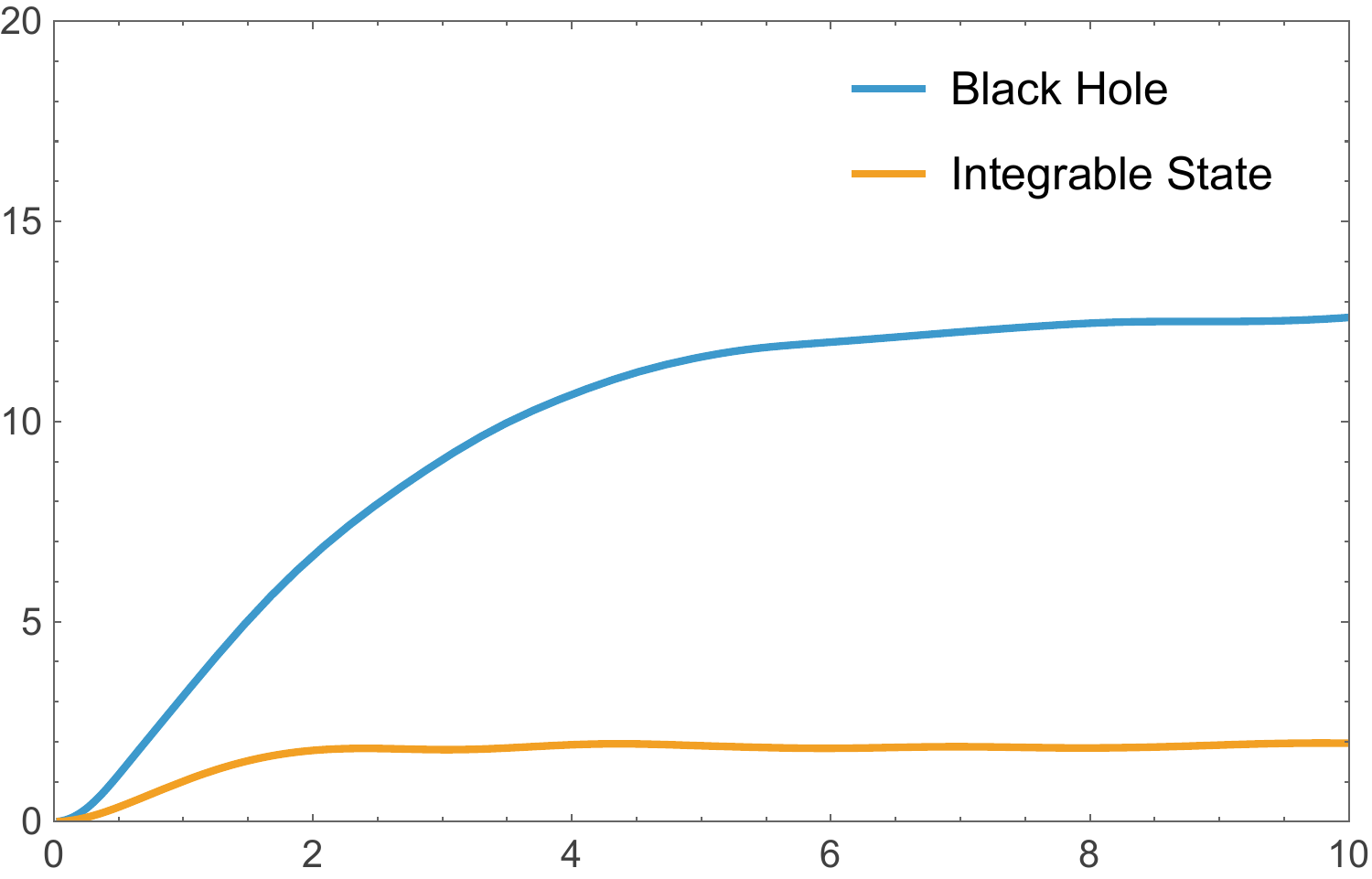}}
\caption{\label{CKave} The time averaged Krylov complexity for respective the integrable and the black hole state for $L=16$.}
\end{figure}

In fig.~\ref{CKave}  we show the time averaged Krylov complexity of  
the black hole state compared to that of the integrable two-site singlet, for $L=16$. We remark that the part of the spin chain Hilbert space available for the black hole state amounts to 69 parity even states out of the 1430 states with non-crossing 
pairings  for $L=16$.
While the limitation on  the available Hilbert space does not allow us to make strong conclusions and calls for more 
investigations,
we do observe that the complexity of the black-hole state grows more rapidly and saturates and a higher value compared to
the integrable state. The oscillations, not shown in the plot are also more pronounced for the integrable state.   
\section{Thermal One-point Functions \label{thermal}}

We finally return to our original motivation of studying thermal one-point functions in AdS/CFT. We consider the full set of scalar operators in the $\mathcal{N}=4$ super-Yang-Mills theory:
\begin{equation}
 \mathcal{O}=\Psi ^{i_1\ldots i_L}\mathop{\mathrm{tr}}\Phi _{i_1}\ldots \Phi _{i_L},
\end{equation}
where $i_1,\ldots,i_L \in \{1,\ldots,6\}$. The one-loop dilatation operator acting on these states can be identified with the Hamiltonian of an $SO(6)$ spin chain: $H=\sum_{\ell}h_{\ell,\ell+1}$ with $h=2-2P+K$, where $P$ and $K$ are the permutation and  trace operators \cite{Minahan:2002ve}, and this Hamiltonian is integrable.

To compute the one-point functions in the thermal ensemble, to the leading order in perturbation theory, we need 
to contract identical fields pairwise with the thermal propagator 
\begin{equation}
 D_\beta (x)=\frac{1}{4\pi ^2}\sum_{n=-\infty }^{+\infty }\frac{1}{\left(t+n\beta \right)^2+\mathbf{x}^2}\,,
\end{equation}
evaluated at coincident points.
Thermal effects are encoded in the difference $D_\beta -D_\infty $, which amounts in omitting the $n=0$ term. Each bubble then contributes
\begin{equation}
 \frac{1}{4\pi ^2}\sum_{n\neq 0}^{ }\frac{1}{n^2\beta^2}=\frac{1}{12\beta ^2}\,.
\end{equation}
The full planar
contribution to the one-point function is accounted for by
the following spin chain overlap:
\begin{equation}
 \left\langle \mathcal{O}\right\rangle=\frac{1}{\sqrt{L}}\left(\frac{\pi T}{\sqrt{3}}\right)^L\frac{\left\langle {\rm BH} \right.\!\left| \Psi \right\rangle}{\left\langle \Psi \right.\!\left| \Psi \right\rangle^{\frac{1}{2}}}\, ,
\end{equation}
where the prefactor accounts for the  unit normalization of the two-point function.
The boundary state is the direct counterpart of the black-hole state that we have studied in the Heisenberg model:
\begin{equation}
 {\rm BH}^{i_1\ldots i_L}=\lim_{N\rightarrow \infty }\int_{}^{}d\phi ^i\,\,{\rm e}\,^{-\,\mathop{\mathrm{tr}}\phi ^i\phi ^i}\mathop{\mathrm{tr}}\phi ^{i_1}\ldots \phi ^{i_L}.
\end{equation}
The integration here is over six Hermitian $N\times N$ matrices with the usual numerical entries. The state can be also constructed from a recursion relation:
\begin{equation}
 \left|{\rm BH}_L\right\rangle=\sum_{n=1}^{L/2}\sum_{i=1 }^{6}
\left|i \right\rangle
 \otimes\left|{\rm BH}_{2n-2}\right\rangle
 \otimes\left|i \right\rangle
 \otimes\left|{\rm BH}_{L-2n}\right\rangle.
\end{equation}
This state shares many features with its $SU(2)$ counterpart. In particular, it breaks integrability: one can check that $Q_3\left|{\rm BH}\right\rangle\neq 0$ for $Q_3=i\sum_{\ell}^{}[h_{\ell,\ell+1},h_{\ell+1,\ell+2}]$. In contradistinction to the Heisenberg model, the planar contractions do not constitute a full basis of singlets for the $SO(6)$ spin chain, as the Rumer identity  does not hold for $SO(6)$ spins.

\section{conclusion\label{Conclusion}}

We have introduced the concept of a black-hole state in a quantum spin chain motivated by  thermal
correlators in holography.  
As an initial step
we concentrated on the black-hole states in the Heisenberg spin chain and provided evidence that they thermalize at
infinite temperature. In addition, we demonstrated that their entanglement entropy grows logarithmically with length as is
typical for  critical, gapless systems. It would be important to develop a systematic explanation for this observation.          
\newline \indent
The black-hole state which arises directly from thermal correlators in holography is also not  integrable. We leave a comprehensive investigation of its  thermal and complexity properties to future work.
  Besides giving information about the infall of a particle into the black hole~\cite{Grinberg:2020fdj} and thermal phase transitions~\cite{Witten:1998zw}, 
thermal one-point functions  have an important role to play as input to the thermal
bootstrap program~\cite{Barrat:2025twb,Alkalaev:2024jxh,Buric:2024kxo}, not least in relation to the
investigation of two-point
correlators as probes sensitive to the internal structure of  the  
space-time singularity~\cite{Ceplak:2024bja,Ceplak:2025dds,Buric:2025fye}.

\vspace*{0.3cm}

\paragraph{Acknowledgements}
We would like to thank Pawel Caputa for very useful discussions. CK was supported by Villum Fonden via the Villum Investigator grant 73742. KZ was supported by VR grant 2021-04578. CK would like to thank the Isaac Newton Institute for Mathematical Sciences, Cambridge, for support and hospitality during the programme Diving Deeper into Defects: On the Intersection of Field Theory, Quantum Matter, and Mathematics, where a part of the work on this paper was undertaken. This work was supported by EPSRC grant EP/Z000580/1.

\bibliography{refs-thermal}

\end{document}